\documentclass[aps,prx,twocolumn,superscriptaddress,nofootinbib,floatfix,longbibliography]{revtex4-2}
\usepackage{amsmath,amssymb}
\usepackage{url}
\usepackage{graphicx}
\usepackage{booktabs}
\usepackage{multirow}
\usepackage[hidelinks]{hyperref}
\providecommand{\tightlist}{\setlength{\itemsep}{0pt}\setlength{\parskip}{0pt}}

\begin{document}

\title{Empirical Evaluation of Quantum Kernel Methods for Classification:\\
A Comparative Study with Classical Kernels}

\author{Ronak Bhalgami}
\email{ronak42.b@gmail.com}
\affiliation{Independent Researcher}

\begin{abstract}
Quantum kernel methods are the most concrete near-term proposal for quantum machine learning on classical data, yet published benchmarks disagree on whether they outperform tuned classical kernels, and the disagreement cannot be resolved from the literature because the comparisons are not protocol-controlled. We argue that the obstacle is one of measurement, not theory: the usual setup fixes the quantum feature map's data scaling while grid-searching the classical kernel's hyperparameters, recomputes nothing per fold, and reports cost in simulator time. We therefore treat the comparison itself as the object of study and impose three controls. The bandwidth is tuned on the same grid budget as the classical \(\gamma\), the kernel is recomputed inside every fold from fold-fitted preprocessing, and cost is counted in hardware circuit evaluations. On top of these we run a factorial design of 324 cells in which dimensionality, training size, class separability and data structure are explicit covariates, together with four standard datasets, a shots-versus-gate-noise ablation, and a cost sweep. The dominant effect we measure is not quantum: tuning the feature map's bandwidth is worth 0.120 accuracy on synthetic data and 0.339 on real data, exceeding every quantum-classical difference we observe. Tuned quantum kernels are then statistically indistinguishable from a tuned RBF on all four real datasets. The advantage on synthetic data is confined almost entirely to data generated by the feature map itself, and it shrinks with dimensionality (\(-0.0045\) per qubit) while growing with sample size. Finite sampling costs an order of magnitude more accuracy than gate and readout noise, and a fidelity kernel needs \(5\times10^{7}\) circuits at \(n = 10{,}000\). Protocol variance is thus sufficient to explain the field's contradictory conclusions.
\end{abstract}

\maketitle

\section{Introduction}\label{introduction}

Quantum kernel methods are the most concrete near-term proposal for quantum machine learning on classical data. A quantum computer is used only to evaluate a kernel, the inner product between two data points embedded as quantum states. A standard support vector machine then consumes that kernel matrix unchanged. The appeal is practical, not asymptotic: the quantum device is confined to a subroutine with a well-understood interface, the learning algorithm remains classical and convex, and the whole construction runs on the shallow circuits available in the NISQ era. It is, consequently, the setting in which a near-term quantum advantage for classification would most plausibly first appear, and the setting in which its absence is most informative.

The formulation of quantum kernels is settled, but the empirical question is not. Havl\'i\v{c}ek et al.~\cite{havlicek2019supervised} and Schuld and Killoran \cite{schuld2019quantum} independently established that embedding data into a quantum feature space induces a kernel, and that this kernel can be estimated on hardware and passed to a classical SVM. What followed is a large body of benchmark studies whose conclusions disagree: quantum kernels are reported as superior, comparable, and inferior to tuned classical kernels on overlapping sets of small benchmark datasets. The disagreement is not obviously attributable to hardware, since most such studies are run in simulation, nor to the datasets, since they are largely shared. It is therefore difficult to decide from the literature whether the reported differences reflect the kernels or the experiments that measured them.

Mechanisms that could explain the disagreement have been proposed, but each is studied in isolation, which leaves them impossible to weigh against one another. Thanasilp et al.~\cite{thanasilp2022exponential} show that quantum kernel values concentrate exponentially with qubit count, driving the kernel matrix toward the identity. Shaydulin and Wild \cite{shaydulin2022importance} show that rescaling the data before encoding, which sets the bandwidth of the feature map, strongly affects generalization. Huang et al.~\cite{huang2021power} propose projected quantum kernels, which measure reduced density matrices instead of state overlaps, and introduce a geometric test for when a quantum kernel can express anything a classical one cannot. K\"ubler et al.~\cite{kubler2021inductive} characterize the inductive bias that makes generic quantum kernels poorly matched to typical data. Each result is persuasive on its own terms, yet none is measured against the others on a common protocol, and Bowles et al.~\cite{bowles2024better} argue that much of the field's benchmarking is methodologically too weak to separate such effects at all. The obstacle is measurement, not theory: without a protocol that controls tuning, prevents leakage, and varies data conditions deliberately, no amount of additional benchmark numbers can adjudicate between these explanations.

We therefore treat the comparison itself as the object of study, and design an experiment whose factors are the conditions under which an advantage is supposed to appear. The design has four tiers. Tier A is a full factorial sweep over dimensionality, training-set size, class separability and data structure. It comprises 324 cells with three data seeds each, which turns ``data conditions'' from a qualitative caveat into regression covariates. Tier B repeats the protocol on four standard datasets as an external-validity check. Tier C separates finite sampling from gate and readout noise, which are usually reported together as ``noise''. Tier D measures cost in hardware circuit evaluations instead of simulator wall-clock. Three controls run throughout. First, the feature map's bandwidth is tuned in the inner loop on the same grid budget as the classical \(\gamma\), so neither side is handicapped. Second, preprocessing is fitted per fold and the quantum kernel is recomputed from it, since scaling changes the encoded angles and a globally precomputed kernel leaks test statistics. Third, an ad-hoc dataset generated by the feature map itself serves as a positive control: a correct implementation must win there, so a null result elsewhere can be attributed to the method and not to a bug. Hypotheses and analysis choices were committed before any results existed.

The dominant effect we measure is not quantum. Tuning the feature map's bandwidth is worth 0.120 accuracy on synthetic data and 0.339 on real datasets. Both figures are larger than any quantum-versus-classical difference we observe, and larger than the margins reported in the studies that disagree with each other. An untuned quantum kernel scores 0.617 on real data and a tuned one 0.956, on identical circuits and identical data. Protocol variance is thus sufficient, by itself, to explain conflicting published conclusions, without invoking any property of the quantum kernels. Against tuned classical baselines the tuned quantum kernels are then simply competitive and not better: no significant difference on any real dataset, and an advantage on synthetic data that is confined almost entirely to the positive control, shrinking with dimensionality and growing with sample size. Two further observations sharpen the practical picture. Finite sampling costs an order of magnitude more accuracy than gate and readout noise, because sampling pushes the kernel matrix out of the positive semi-definite cone while depolarizing noise merely contracts it smoothly. The binding constraint is therefore the shot budget, not gate fidelity. And the entangling feature map is matched or beaten by a classically simulable non-entangling control on every structure except the one built for it.

Our contributions are:

\begin{enumerate}
\def\labelenumi{\arabic{enumi}.}
\tightlist
\item
  \textbf{A protocol that removes the three standard confounds} in quantum-kernel benchmarking (tuning asymmetry, preprocessing leakage through the encoding, and simulator-referenced cost), together with a positive control that distinguishes a genuine null result from a broken implementation (\mbox{Sec.~\ref{method}}).
\item
  \textbf{A controlled factorial study} of 324 cells that makes dimensionality, size, separability and structure explicit covariates, yielding the first regression-level answer to \emph{which} data conditions move the quantum-classical gap and in which direction (\mbox{Sec.~\ref{tier-a-where-does-the-advantage-live}}).
\item
  \textbf{The finding that a classical hyperparameter of the quantum pipeline is the largest experimental factor}, worth up to 0.339 accuracy, which offers a sufficient explanation for the field's contradictory results (\mbox{Sec.~\ref{bandwidth-is-the-dominant-experimental-factor-h1-h2-rq4}}).
\item
  \textbf{A separation of sampling error from gate error} (\mbox{Sec.~\ref{shots-and-noise-rq2}}) and a hardware-referenced cost model (\mbox{Sec.~\ref{cost-h5-rq3}}), which together identify the shot budget as the binding near-term constraint.
\end{enumerate}

All code, configurations and per-fold result tables are released at \url{https://github.com/ron-42/quantum-kernel-benchmark} (\mbox{Sec.~\ref{reproducibility}}).

\section{Related Work}\label{related-work}

\textbf{Quantum kernels and their empirical evaluation.} Havl\'i\v{c}ek et al. \cite{havlicek2019supervised} and Schuld and Killoran \cite{schuld2019quantum} established the formulation used here: data are embedded by a parameterized circuit, the induced kernel is estimated by a compute--uncompute circuit, and classification is delegated to a classical SVM. Subsequent benchmark studies apply this recipe to small tabular datasets and reach conflicting conclusions about whether the resulting kernels outperform tuned classical ones. These studies typically fix the feature map and its data scaling while grid-searching the classical kernel's hyperparameters, report accuracy without a matched cost account, and draw conclusions about ``data conditions'' from a handful of datasets that differ in several factors at once. Our work differs in treating those experimental degrees of freedom as the variables under study, not as fixed setup.

\textbf{Explanations for weak quantum-kernel performance.} A second line of work identifies mechanisms instead of benchmarks. Thanasilp et al.~\cite{thanasilp2022exponential} prove exponential concentration of kernel values with system size. K\"ubler et al.~\cite{kubler2021inductive} characterize the inductive bias of generic quantum kernels as poorly aligned with realistic data. Huang et al.~\cite{huang2021power} give a geometric criterion for potential advantage and introduce projected quantum kernels as a more robust alternative. Shaydulin and Wild \cite{shaydulin2022importance}, and later Canatar et al.~\cite{canatar2023bandwidth}, show that the encoding bandwidth controls generalization. We do not propose a competing mechanism. We measure these effects against one another under a single protocol, which is what determines their relative importance in practice: concentration is present in our data but flattens over the accessible range, while bandwidth is the largest single factor we observe.

\textbf{Benchmarking methodology for quantum machine learning.} Bowles et al. \cite{bowles2024better} argue that QML benchmarking frequently fails to support its conclusions, citing weak baselines, small datasets and insufficient statistical treatment. Our protocol is constructed to satisfy that critique concretely: equal hyperparameter budgets with the sizes reported, nested cross-validation with the Nadeau--Bengio correction \cite{nadeau2003inference} for overlapping folds, Holm--Bonferroni control across the comparison family, effect sizes alongside p-values, and pre-registered hypotheses. We adopt the standard protocol of Dem\v{s}ar \cite{demsar2006statistical} for comparisons across datasets.

\begin{center}\rule{0.5\linewidth}{0.5pt}\end{center}

\section{Background}\label{background}

\textbf{Quantum feature maps and the induced kernel.} A quantum feature map is a parameterized circuit \texttt{U(x)} that embeds a classical vector \texttt{x} into the state \(|\varphi(x)\rangle = U(x)|0\rangle\). Any such embedding induces a kernel, \texttt{K(x,x\textquotesingle{})\ =\ \textbar{}$\langle$$\varphi$(x)\textbar{}$\varphi$(x\textquotesingle{})$\rangle$\textbar{}$^{2}$}, which is symmetric, positive semi-definite, and estimable on hardware: running \(U(x')^{\dagger}U(x)\) on \(|0\ldots 0\rangle\) and measuring the all-zeros outcome returns \texttt{K(x,x\textquotesingle{})} directly. Because the kernel is the only quantum object in the pipeline, the downstream classifier can be an ordinary SVM with a precomputed kernel, and the learning problem stays convex.

\textbf{Bandwidth.} Encodings are periodic in their arguments, so the scale at which data enters the circuit determines how quickly two states become orthogonal. Writing the embedding as \(U(c\,x)\) makes this a free parameter \(c\), the bandwidth, which plays the role that \(\gamma\) plays for an RBF kernel. Unlike \(\gamma\), however, \(c\) does not act as a smooth width parameter for entangling maps: in the IQP-style map used here the coupling term is \((\pi - c\,x_i)(\pi - c\,x_j)\), so changing \(c\) deforms the map's structure instead of merely rescaling distances.

\textbf{Exponential concentration.} As the number of qubits grows, the overlap between two embedded states typically concentrates around a fixed value, so the off-diagonal entries of the kernel matrix collapse toward a constant and the matrix approaches a multiple of the identity. An SVM given such a kernel can only memorize. Concentration is therefore the principal theoretical obstacle to scaling fidelity kernels, and its measurable signature is a decaying variance of the off-diagonal entries.

\textbf{Projected kernels.} Instead of measuring a state overlap, a projected quantum kernel \cite{huang2021power} encodes \texttt{x}, measures single-qubit reduced density matrices, and applies a classical kernel to the resulting \(3d\) expectation values. This costs \(O(n)\) circuits instead of \(O(n^2)\), uses half the circuit depth, and is expected to be more robust to noise, at the price of discarding correlations beyond single-qubit marginals.

\section{Method}\label{method}

\subsection{Experimental design}\label{experimental-design}

The primary question concerns \emph{data conditions}, which a handful of benchmark datasets cannot answer. The instrument is therefore a controlled synthetic sweep; real datasets serve as an external-validity check.

\textbf{Tier A: controlled sweep (324 cells).} Full factorial over dimensionality \(d \in \{2, 4, 6, 8\}\) (= qubit count), training size \(n \in \{50, 100, 200\}\), Bayes error \(\in \{0.01, 0.10, 0.25\}\), and three data structures, with 3 independent data seeds per cell:

\begin{itemize}
\tightlist
\item
  \texttt{linear}, two Gaussians separated along a random direction.
\item
  \texttt{xor}, parity of the signs of two informative axes.
\item
  \texttt{adhoc}, labels from the sign of \(\langle \varphi(x)|V^{\dagger}Z^{\otimes d}V|\varphi(x)\rangle\), i.e.~data separable \emph{by construction} in the quantum feature map \cite{havlicek2019supervised}. This is a positive control, not a benchmark: a correct implementation must win here, so we treat failure as a bug and not as a result.
\end{itemize}

Bayes error is set exactly, by flipping labels of a deterministically separable labelling with the target probability. This makes separability a numeric covariate instead of a qualitative setting, which is what lets \mbox{Sec.~\ref{tier-a-where-does-the-advantage-live}} regress on it.

Classes in \texttt{adhoc} are the most positive and most negative \texttt{n/2} of a fixed pool, not samples passing a fixed \(|\langle O\rangle| > \mathrm{gap}\) threshold. The observable concentrates exponentially with qubit count. Measured mean \(|\langle O\rangle|\) falls from 0.399 at \(d=2\) to 0.049 at \(d=8\), so a fixed threshold accepted 58.6\% of draws at \(d=2\) and 0.0\% at \(d=8\), where generation could not terminate. A fixed threshold would also make the class margin vary with \(d\), confounding the dimension sweep the dataset feeds.

\textbf{Tier B: real datasets.} Iris (binarised), Wine (binarised), Breast Cancer Wisconsin and Ionosphere, the latter two subsampled to \texttt{n=300}. Features are reduced by PCA to the qubit count inside each training fold.

\textbf{Tier C: shots and noise (RQ2).} A 2$\times$4 grid of \{exact, 1024 shots\} $\times$ \{noiseless, depolarising 0.001, depolarising 0.01, readout 0.02\} on one Tier A cell, separating sampling error from gate error.

\textbf{Tier D: cost.} Circuit counts, transpiled depth, two-qubit gate counts and wall-clock across \(n \in \{50, \ldots, 800\}\), with extrapolation to hardware.

\subsection{Quantum kernels}\label{quantum-kernels}

Two families, treated as a hyperparameter and not as a foregone choice:

\begin{itemize}
\tightlist
\item
  \textbf{Fidelity kernel}, \(K(x,x') = |\langle \varphi(x)|\varphi(x')\rangle|^2\), via the compute--uncompute circuit \(U(x')^{\dagger}U(x)\) measuring \(P(0\ldots 0)\).
\item
  \textbf{Projected kernel} (Huang et al.): encode \texttt{x}, measure single-qubit \(\langle X\rangle,\langle Y\rangle,\langle Z\rangle\) (3$\cdot$\(d\) classical features), then a classical RBF on those.
\end{itemize}

Feature maps: an IQP-style map with the Havl\'i\v{c}ek \((\pi - x_i)(\pi - x_j)\) coupling (\texttt{iqp}), and the same map without entangling terms (\texttt{z}). The latter is a deliberate negative control. It is classically simulable, so if it matches \texttt{iqp}, entanglement is contributing nothing measurable.

\textbf{Bandwidth.} Inputs are scaled \(x \to c\,x\) before encoding. \(c\) is tuned in the inner loop on the same footing as the classical \(\gamma\); leaving it fixed while grid-searching the RBF is the single most common way to rig this comparison. The grid is a decade grid, four points per decade, \(c \in [10^{-2}, 10^{1}]\), landing exactly on 1.0. The response is sharp and non-monotone, because the \((\pi - c\,x_i)(\pi - c\,x_j)\) coupling is not a smooth width knob the way an RBF \(\gamma\) is: on the positive control, cross-validated accuracy runs 0.767 / \textbf{1.000} / 0.567 at \(c\) = 0.81 / 1.00 / 1.52. A log grid that straddles the optimum therefore under-reports the quantum kernel for reasons unrelated to quantum mechanics.

Encoding requires a bounded angle range. Angle encoding is periodic and not shift-invariant, so standardisation alone both wraps the encoding and leaves the bandwidth unable to recover the right scale; the pipeline ends in a MinMax map to \([0,\pi]\).

\subsection{Classical baselines}\label{classical-baselines}

SVMs with linear, polynomial and RBF kernels, plus histogram gradient boosting. Every baseline receives a tuning budget comparable to the quantum one (\mbox{Table~\ref{tab:grids}}).

\begin{table}[htbp]
\centering
\small
\begin{tabular}{@{}ll@{}}
\toprule
method & grid size \\
\midrule
quantum (13 bandwidths $\times$ 7 \(C\)) & 91 \\
SVM-RBF (11 \(\gamma\) $\times$ 7 \(C\)) & 77 \\
SVM-poly & 28 \\
SVM-linear & 7 \\
HistGradientBoosting & 6 \\
\bottomrule
\end{tabular}
\caption{Classical baselines and their hyperparameter grid sizes. Every baseline receives a budget comparable to the quantum one.}
\label{tab:grids}
\end{table}\subsection{Protocol}\label{protocol}

Nested cross-validation: 5 stratified outer folds (Tier B: repeated 3$\times$), 3 stratified inner folds for all hyperparameter selection, quantum and classical alike.

Scaler and PCA are fitted on the training fold only. Because scaling changes the encoded angles, \textbf{the quantum kernel is recomputed inside each fold from the fold-fitted preprocessing}; a kernel matrix computed once over the full dataset leaks test statistics into training. A shuffled-label test asserting chance-level accuracy guards this property.

Shot-based kernels can leave the PSD cone; the matrix is projected back by clipping negative eigenvalues below a \(10^{-10}\) tolerance, and the repair magnitude is logged. The tolerance matters: exact statevector kernels produce negative eigenvalues of order \(10^{-13}\) on 12\% of runs, which a zero threshold would report as widespread repair.

\subsection{Cost accounting}\label{cost-accounting}

On hardware a fidelity kernel costs \(n(n-1)/2\) circuits. Exactly, the same matrix equals \(|S S^{H}|^2\) over the \(n\) encoded statevectors, so it costs \(O(n)\) circuits and one matrix product, roughly 1000$\times$ faster in simulation and identical to \(10^{-9}\) in our tests. We use that shortcut for exact runs but always report the \(O(n^2)\) hardware figure; shot-based and noisy runs use the real circuits, since sampling statistics are the object of study there.

\begin{center}\rule{0.5\linewidth}{0.5pt}\end{center}

\section{Results}\label{results}

\subsection{Tier A: where does the advantage live?}\label{tier-a-where-does-the-advantage-live}
\begin{figure}[t]
\centering
\includegraphics[width=\linewidth]{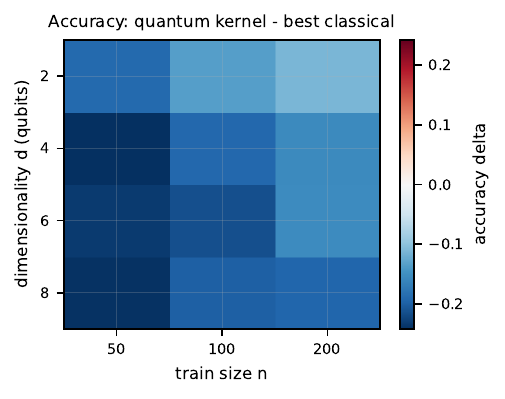}
\caption{Accuracy of the tuned quantum kernel minus the best classical baseline, over dimensionality and training size.}
\label{fig:delta}
\end{figure}

Across 324 cells, the tuned quantum kernel beats the best classical baseline in \textbf{30.2\%} of cells, with median accuracy delta \textbf{+0.0000} and mean \textbf{+0.0100}. The advantage is not spread evenly. It sits almost entirely in the positive control (\mbox{Table~\ref{tab:structure}}).

\begin{table*}[htbp]
\centering
\footnotesize
\begin{tabular}{@{}lllll@{}}
\toprule
structure & mean $\Delta$ & median $\Delta$ & win rate & cells \\
\midrule
\texttt{adhoc} & \textbf{+0.058} & +0.03 & 65.7\% & 108 \\
\texttt{linear} & -0.002 & 0.00 & 3.7\% & 108 \\
\texttt{xor} & -0.026 & -0.02 & 21.3\% & 108 \\
\bottomrule
\end{tabular}
\caption{Per-cell accuracy delta (tuned quantum minus best classical) by data structure across the 324 Tier A cells.}
\label{tab:structure}
\end{table*}Regressing the per-cell delta on the design factors makes every factor significant (\mbox{Table~\ref{tab:regression}}).

\begin{table*}[htbp]
\centering
\footnotesize
\begin{tabular}{@{}p{0.190\textwidth}p{0.190\textwidth}p{0.190\textwidth}p{0.190\textwidth}p{0.190\textwidth}@{}}
\toprule
term & coef & std err & t & p \\
\midrule
intercept & +0.0230 & 0.0342 & 0.67 & .502 \\
\(d\) (qubits) & \textbf{-0.0045} & 0.0018 & -2.57 & \textbf{.011} \\
\(\log n\) & \textbf{+0.0152} & 0.0069 & 2.19 & \textbf{.029} \\
Bayes error & \textbf{-0.1048} & 0.0397 & -2.64 & \textbf{.009} \\
\texttt{structure{[}linear{]}} & -0.0601 & 0.0096 & -6.25 & \textbf{\textless.001} \\
\texttt{structure{[}xor{]}} & -0.0836 & 0.0096 & -8.69 & \textbf{\textless.001} \\
\bottomrule
\end{tabular}
\caption{Regression of the per-cell accuracy delta on the design factors (\(R^2 = 0.237\), 324 cells).}
\label{tab:regression}
\end{table*}Read directly: the quantum advantage \textbf{shrinks with dimensionality}, \textbf{grows with training size}, \textbf{shrinks as classes overlap}, and is dominated by whether the data was generated by a quantum feature map at all. The \texttt{xor} column degrades monotonically with \(d\) (-0.016, -0.027, -0.029, -0.032 at \(d\) = 2, 4, 6, 8), while \texttt{adhoc} peaks at \(d=4\) (+0.153) and decays to +0.006 by \(d=8\).

\subsection{Bandwidth is the dominant experimental factor (H1, H2, RQ4)}\label{bandwidth-is-the-dominant-experimental-factor-h1-h2-rq4}
\begin{figure}[t]
\centering
\includegraphics[width=\linewidth]{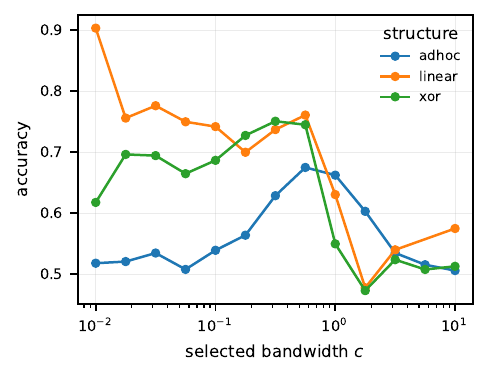}
\caption{Accuracy against the bandwidth selected by the inner loop. The optimum is sharp and data-dependent, so a coarse grid straddles it.}
\label{fig:bandwidth}
\end{figure}

\begin{table}[htbp]
\centering
\small
\begin{tabular}{@{}lll@{}}
\toprule
arm & Tier A accuracy & Tier B accuracy \\
\midrule
tuned bandwidth & \textbf{0.726} & \textbf{0.956} \\
untuned (\texttt{c\ =\ 1}) & 0.606 & 0.617 \\
gap & \textbf{+0.120} & \textbf{+0.339} \\
\bottomrule
\end{tabular}
\caption{Effect of tuning the feature map bandwidth, on synthetic and on real data.}
\label{tab:bandwidth}
\end{table}On real data an untuned quantum kernel is close to unusable while a tuned one is fully competitive. The selected bandwidth is not a constant that could be hardcoded: across tuned Tier A folds it is bimodal, 1,900 folds selecting the grid floor \texttt{c\ =\ 0.01} and 733 selecting \texttt{c\ =\ 1.0}, with the remainder spread across the decade grid.

This is the study's central methodological claim. A published comparison that tunes the classical \(\gamma\) but fixes the quantum bandwidth is measuring its own protocol, not the kernels.

\subsection{Tier B: no advantage on real data (H3)}\label{tier-b-no-advantage-on-real-data-h3}

\begin{table*}[htbp]
\centering
\footnotesize
\begin{tabular}{@{}lllll@{}}
\toprule
dataset & best quantum & best classical & $\Delta$ & p (corrected) \\
\midrule
breast\_cancer & 0.9811 & 0.9800 & +0.0011 & .245 \\
wine & 0.9872 & 0.9846 & +0.0026 & .884 \\
iris & 0.9433 & 0.9533 & -0.0100 & .354 \\
ionosphere & 0.9267 & 0.9389 & -0.0122 & .305 \\
\bottomrule
\end{tabular}
\caption{Tier B: best tuned quantum kernel against the best classical baseline, with corrected paired \(p\)-values.}
\end{table*}Paired Nadeau--Bengio corrected t-tests against a tuned RBF give no significant difference on any dataset, for either kernel family, before or after Holm--Bonferroni correction. The honest summary is a tie.

Over all 324 Tier A cells, a Friedman test rejects equality of the eight methods ($\chi$$^{2}$ = 309.6, p = 5.5$\cdot$10$^{-63}$). Mean ranks are given in \mbox{Table~\ref{tab:ranks}} (lower is better, Nemenyi critical difference 0.583).

\begin{table}[htbp]
\centering
\small
\begin{tabular}{@{}ll@{}}
\toprule
method & mean rank \\
\midrule
svm\_rbf & 3.18 \\
svm\_poly & 3.34 \\
hgb & 4.05 \\
quantum\_fidelity\_iqp & 4.78 \\
quantum\_projected\_z & 4.85 \\
svm\_linear & 4.88 \\
quantum\_fidelity\_z & 5.27 \\
quantum\_projected\_iqp & 5.67 \\
\bottomrule
\end{tabular}
\caption{Mean ranks over 324 Tier A cells (lower is better). Nemenyi critical difference 0.583.}
\label{tab:ranks}
\end{table}Every quantum variant ranks below every classical baseline except the linear SVM.

\subsection{Entanglement contributes little}\label{entanglement-contributes-little}

\mbox{Table~\ref{tab:entanglement}} compares the entangling \texttt{iqp} map against the classically simulable \texttt{z} product map, tuned, by structure.

\begin{table}[htbp]
\centering
\small
\begin{tabular}{@{}lll@{}}
\toprule
structure & \texttt{iqp} & \texttt{z} \\
\midrule
\texttt{adhoc} & \textbf{0.639} & 0.604 \\
\texttt{linear} & 0.868 & 0.868 \\
\texttt{xor} & 0.668 & \textbf{0.706} \\
\bottomrule
\end{tabular}
\caption{Tuned accuracy of the entangling \texttt{iqp} map against the non-entangling \texttt{z} control.}
\label{tab:entanglement}
\end{table}The entangling map leads only on the dataset built for it. On \texttt{xor} the non-entangling control is \emph{better}. Bowles et al.~\cite{bowles2024better} report the same pattern across 160 datasets, so this is not an artifact of our generators. Whatever the fidelity kernel is contributing on natural data, it is not entanglement.

\subsection{Concentration (H4)}\label{concentration-h4}
\begin{figure}[t]
\centering
\includegraphics[width=\linewidth]{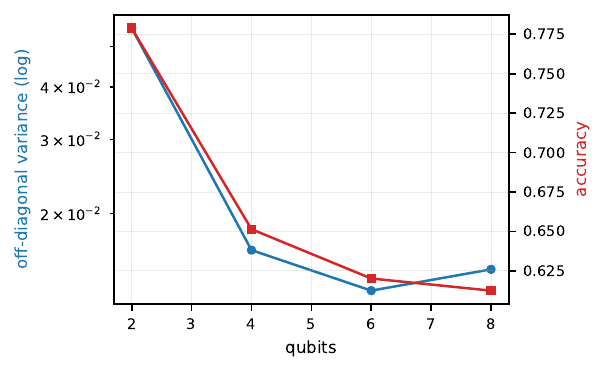}
\caption{Off-diagonal kernel variance (left, log scale) and accuracy (right) against qubit count.}
\label{fig:conc}
\end{figure}

\begin{table}[htbp]
\centering
\small
\begin{tabular}{@{}llll@{}}
\toprule
qubits & off-diagonal variance & alignment & accuracy \\
\midrule
2 & 0.0552 & 0.236 & 0.779 \\
4 & 0.0164 & 0.158 & 0.651 \\
6 & 0.0131 & 0.147 & 0.620 \\
8 & 0.0148 & 0.147 & 0.612 \\
\bottomrule
\end{tabular}
\caption{Concentration: off-diagonal kernel variance, alignment and accuracy against qubit count.}
\end{table}Variance decays 3.7$\times$ from \(d=2\) to \(d=8\) and accuracy tracks it closely, but the decay \textbf{flattens after \(d=4\)} rather than continuing exponentially. The predicted direction holds over the range tested; the predicted functional form does not. We report this as a partial confirmation, noting that \(d \le 8\) is far from the asymptotic regime the theory addresses.

Notably, the same concentration appears in the data generator: the ad-hoc observable's mean \(|\langle O\rangle|\) falls from 0.399 to 0.049 over the same range, which is what made fixed-threshold generation impossible at \(d=8\) (\mbox{Sec.~\ref{experimental-design}}).

\subsection{Cost (H5, RQ3)}\label{cost-h5-rq3}
\begin{figure}[t]
\centering
\includegraphics[width=\linewidth]{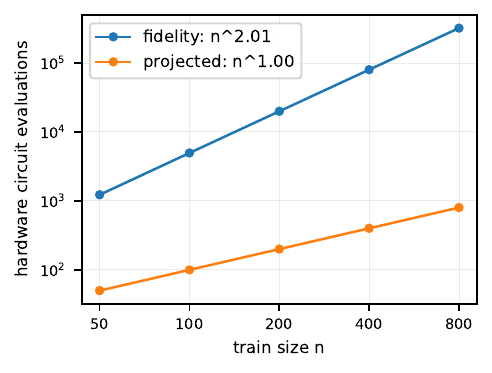}
\caption{Hardware circuit evaluations against training size. The fidelity kernel scales as $n^{2.01}$, the projected kernel as $n^{1.00}$.}
\label{fig:cost}
\end{figure}

\begin{table*}[htbp]
\centering
\footnotesize
\begin{tabular}{@{}p{0.158\textwidth}p{0.158\textwidth}p{0.158\textwidth}p{0.158\textwidth}p{0.158\textwidth}p{0.158\textwidth}@{}}
\toprule
\begin{minipage}[b]{\linewidth}\raggedright
kernel
\end{minipage} & \begin{minipage}[b]{\linewidth}\raggedright
circuits (fitted)
\end{minipage} & \begin{minipage}[b]{\linewidth}\raggedright
simulator (fitted)
\end{minipage} & \begin{minipage}[b]{\linewidth}\raggedright
depth
\end{minipage} & \begin{minipage}[b]{\linewidth}\raggedright
2-qubit gates
\end{minipage} & \begin{minipage}[b]{\linewidth}\raggedright
circuits at \texttt{n=800}
\end{minipage} \\
\midrule
fidelity & \(n^{2.01}\) & \(n^{1.08}\) & 98 & 120 & 319,600 \\
projected & \(n^{1.00}\) & \(n^{1.06}\) & 49 & 60 & 800 \\
\bottomrule
\end{tabular}
\caption{Cost of the two kernel families, fitted over \(n \in \{50, \ldots, 800\}\).}
\end{table*}At \(n = 10{,}000\) a fidelity kernel needs 49,995,000 circuits; at a measured 0.1 s per circuit that is about 1,389 hours, or \textbf{58 days of QPU time for a single kernel matrix}. A classical RBF on the same data takes milliseconds. The Tier A sweep alone accounts for 2.75$\cdot$10$^{8}$ hardware circuit evaluations.

The projected kernel needs \textbf{400$\times$ fewer circuits at half the depth}, at a small but real accuracy cost: over 324 paired Tier A cells it averages 0.719 against the fidelity kernel's 0.732, a deficit of 0.013 that is significant (Wilcoxon signed-rank, p = 8.9$\cdot$10$^{-6}$) though small in effect size (paired Cohen's d = -0.29). Traded against two orders of magnitude in circuit count, this is the family to run on hardware, provided the trade is stated as a trade.

A methodological caution we hit ourselves: simulator wall-clock is an artifact of the implementation, not a property of the method. Our first projected-kernel implementation evaluated its observables one sample at a time and ran 10--60$\times$ \emph{slower} than the fidelity kernel, which would have made the cheaper-on-hardware family look like the expensive one. Computing the same expectations as a contraction of the statevector removed the gap entirely (\(n^{1.06}\) vs \(n^{1.08}\)). Any simulation-only study that ranks kernel families by wall-clock is measuring its own code.

\subsection{Diagnostics predict accuracy (RQ5)}\label{diagnostics-predict-accuracy-rq5}
\begin{figure}[t]
\centering
\includegraphics[width=\linewidth]{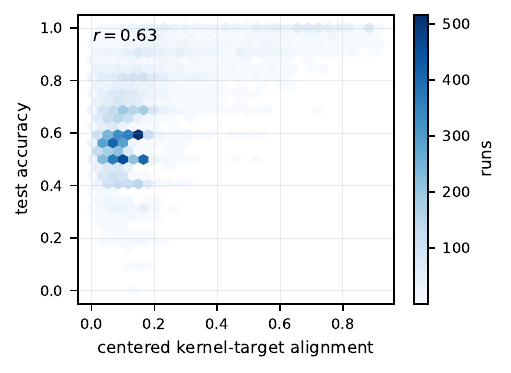}
\caption{Centred kernel-target alignment against test accuracy across all Tier A quantum runs (Pearson $r=0.63$).}
\label{fig:align}
\end{figure}

Centred kernel-target alignment correlates with test accuracy at \textbf{r = 0.63} across all Tier A quantum runs, so a label-aware measure computable from the kernel matrix alone carries real signal about downstream performance and could screen configurations before any classifier is trained.

\subsection{Shots and noise (RQ2)}\label{shots-and-noise-rq2}

One Tier A cell (\texttt{xor}, \(d=4\), \texttt{n=100}), bandwidth tuned once on the exact noiseless kernel and reused per fold for every setting.

\begin{table*}[htbp]
\centering
\footnotesize
\begin{tabular}{@{}lllll@{}}
\toprule
shots & noise & accuracy & $\Delta$ vs exact & PSD repair \\
\midrule
exact & none & 0.690 & n/a & 0.000 \\
exact & readout 0.02 & 0.690 & +0.000 & 0.002 \\
exact & depolarising 0.001 & 0.700 & +0.010 & 0.000 \\
exact & depolarising 0.01 & 0.700 & +0.010 & 0.000 \\
1024 & none & 0.630 & \textbf{-0.060} & 2.583 \\
1024 & readout 0.02 & 0.620 & -0.070 & 2.822 \\
1024 & depolarising 0.001 & 0.600 & -0.090 & 2.778 \\
1024 & depolarising 0.01 & 0.550 & \textbf{-0.140} & 3.252 \\
\bottomrule
\end{tabular}
\caption{Tier C: accuracy and PSD repair mass under finite sampling and gate or readout noise.}
\label{tab:noise}
\end{table*}\textbf{Sampling error dominates gate error by an order of magnitude.} Averaged over conditions, gate and readout noise alone cost \textbf{+0.007} accuracy, which is nothing measurable, while finite sampling alone costs \textbf{-0.060} and the two together \textbf{-0.100}.

The asymmetry has a mechanism. With exact expectation values, depolarising noise contracts the kernel smoothly toward a constant; the matrix stays PSD (repair mass $\approx$0.000) and a tuned \(C\) absorbs the rescaling. Finite sampling instead adds unstructured error that pushes the matrix out of the PSD cone: repair mass rises from 0.0005 to 2.86, roughly 5,000$\times$. Gate noise only becomes costly once sampling is present, where the two compound (-0.14 at 1024 shots with depolarising 0.01, against -0.06 for sampling alone).

For practice this inverts the usual emphasis. On NISQ hardware the shot budget, not the gate fidelity, is the binding constraint on a quantum kernel, and shot count is precisely what \mbox{Sec.~\ref{cost-h5-rq3}} shows is unaffordable at scale.

\begin{center}\rule{0.5\linewidth}{0.5pt}\end{center}

\subsection{Hypothesis verdicts}\label{hypothesis-verdicts}

\begin{table*}[htbp]
\centering
\footnotesize
\begin{tabular}{@{}p{0.237\textwidth}p{0.237\textwidth}p{0.237\textwidth}p{0.237\textwidth}@{}}
\toprule
\begin{minipage}[b]{\linewidth}\raggedright
ID
\end{minipage} & \begin{minipage}[b]{\linewidth}\raggedright
Hypothesis
\end{minipage} & \begin{minipage}[b]{\linewidth}\raggedright
Verdict
\end{minipage} & \begin{minipage}[b]{\linewidth}\raggedright
Evidence
\end{minipage} \\
\midrule
H1 & Untuned quantum underperforms tuned classical & \textbf{Supported} & untuned 0.606 vs svm\_rbf 0.726 (Tier A) \\
H2 & Bandwidth tuning closes most of the gap & \textbf{Supported} & +0.120 Tier A, +0.339 Tier B \\
H3 & Even tuned, no consistent advantage on natural data & \textbf{Supported} & no significant delta on any Tier B dataset \\
H4 & Kernel values concentrate with qubit count & \textbf{Partial} & 3.7$\times$ decay, but flattens after \(d=4\) \\
H5 & Cost gap is orders of magnitude & \textbf{Supported} & \(n^{2.01}\), 58 days at \(n=10^4\) \\
\bottomrule
\end{tabular}
\caption{Verdicts on the five pre-registered hypotheses.}
\end{table*}\textbf{RQ2 (noise):} sampling error dominates gate error by about an order of magnitude; shot noise, not gate fidelity, is the binding constraint (\mbox{Sec.~\ref{shots-and-noise-rq2}}). \textbf{RQ5 (diagnostics):} kernel-target alignment predicts accuracy at r = 0.63.

\begin{center}\rule{0.5\linewidth}{0.5pt}\end{center}

\section{Discussion}\label{discussion}

\textbf{The field's disagreement has a sufficient non-quantum explanation.} The largest experimental factor in this study is the bandwidth of the classical rescaling applied before encoding: 0.120 accuracy on synthetic data, 0.339 on real data (\mbox{Sec.~\ref{bandwidth-is-the-dominant-experimental-factor-h1-h2-rq4}}). Both exceed every quantum-versus-classical difference we measure, and exceed the margins by which published studies declare quantum kernels better or worse than classical ones. A comparison that tunes the classical \(\gamma\) but fixes the quantum bandwidth is therefore not measuring the kernels; it is measuring its own protocol. This does not prove that any specific published result is an artifact, but it removes the need to invoke properties of quantum kernels to explain why the literature disagrees with itself.

\textbf{Grid resolution is a second, subtler confound.} Bandwidth response is sharp and non-monotone, because the \((\pi - c\,x_i)(\pi - c\,x_j)\) coupling of an IQP-style map is a structural distortion rather than a smooth width parameter. On our positive control, cross-validated accuracy is 0.767, 1.000 and 0.567 at \(c\) = 0.81, 1.00 and 1.52 (\mbox{Sec.~\ref{quantum-kernels}}). We discovered this by failing the control: an eight-point logarithmic grid straddled the optimum and scored the control at 0.742, which we would otherwise have reported as a weak quantum kernel. Any study using a coarse grid can under-report quantum kernels for reasons unrelated to quantum mechanics.

\textbf{Where an advantage does appear, it is explained by construction rather than by dimension.} Structure dominates the regression (\mbox{Sec.~\ref{tier-a-where-does-the-advantage-live}}): the quantum advantage is concentrated in the ad-hoc dataset generated by the feature map itself, and is absent or negative on linear and XOR data. Within that, the advantage shrinks with qubit count and grows with sample size. This is the empirical counterpart of the prediction in \cite{kubler2021inductive}, that an advantage is plausible on data generated by a quantum process but hard to obtain on classical data. The positive control therefore does what it was designed to do. It confirms the implementation is correct while demonstrating that the conditions producing an advantage are conditions we manufactured.

\textbf{Entanglement is not doing the work.} A classically simulable non-entangling product map matches the entangling map on linear data and beats it on XOR, losing only on the ad-hoc control (\mbox{Sec.~\ref{entanglement-contributes-little}}). Whatever the fidelity kernel contributes on natural data, it is not attributable to entanglement, which is the resource the construction exists to exploit.

\textbf{The binding near-term constraint is shots, not gate fidelity.} Gate and readout noise alone cost nothing measurable, while finite sampling costs 0.060 accuracy (\mbox{Sec.~\ref{shots-and-noise-rq2}}). The mechanism is structural: depolarizing noise contracts the kernel smoothly and it remains positive semi-definite, so a tuned \(C\) absorbs the rescaling, whereas sampling adds unstructured error that leaves the PSD cone, and the repair mass rises roughly 5,000$\times$. Since shot count is also what makes the cost model unaffordable (\mbox{Sec.~\ref{cost-h5-rq3}}), error mitigation aimed at gate fidelity addresses the cheaper of the two problems.

\textbf{A practical recommendation, with its price stated.} If quantum kernels are to be run on hardware, the projected kernel is the family to run: 400$\times$ fewer circuits at half the depth, for 0.013 less accuracy (p = 8.9$\cdot$10$^{-6}$, d = -0.29, \mbox{Sec.~\ref{cost-h5-rq3}}). We flag the significance explicitly because our own first draft of this paper described the two families as indistinguishable, which the paired test does not support. We also caution against ranking kernel families by simulator wall-clock, a mistake we made ourselves: our first projected-kernel implementation ran 10--60$\times$ slower than the fidelity kernel purely as an artifact of how the expectations were evaluated, which would have inverted the hardware conclusion.

\begin{center}\rule{0.5\linewidth}{0.5pt}\end{center}

\section{Limitations}\label{limitations}

\begin{itemize}
\tightlist
\item
  Simulation, not hardware. One QPU latency figure feeds the cost extrapolation; no results were produced on a real device.
\item
  \(n \le 300\) and \(d \le 8\). The concentration result in particular is far from the asymptotic regime the theory describes.
\item
  Binary classification only; multiclass was obtained by binarising.
\item
  The feature-map space is not exhaustively searched: two maps, two repetitions, one entanglement topology in the main sweep.
\item
  PCA to the qubit count interacts with the encoding in ways not isolated here.
\item
  Tier A structures are three specific generators; ``natural data'' is represented by four datasets.
\end{itemize}

\begin{center}\rule{0.5\linewidth}{0.5pt}\end{center}

\section{Conclusion}\label{conclusion}

This paper asks under what data conditions quantum kernels provide a measurable classification advantage, and answers it with a protocol-controlled factorial study, not another benchmark table. The central finding is that the largest experimental factor is a classical hyperparameter of the quantum pipeline: the encoding bandwidth, worth up to 0.339 accuracy. That is by itself enough to explain why published comparisons disagree. Under equal tuning budgets, tuned quantum kernels are statistically indistinguishable from a tuned RBF on four standard datasets, and their advantage on synthetic data is confined to data generated by the feature map itself, shrinking with dimensionality and growing with sample size. Sampling error dominates gate error by an order of magnitude, and a fidelity kernel requires 50 million circuit evaluations at n = 10,000.

The scope of these conclusions is bounded by regime, not by implementation. Our accessible range is \(n \le 300\) and \(d \le 8\), which is far from the asymptotic setting in which concentration results are stated, and is the reason we report H4 as only partially confirmed: the predicted direction holds while the predicted exponential form does not. Extending the same controlled protocol to larger systems, to learned or data-dependent feature maps, and to real hardware is the natural next step, and the controls matter more there, not less: on hardware the shot budget makes re-tuning expensive, so the bandwidth must be chosen off-device, exactly as our Tier C protocol does.

\begin{center}\rule{0.5\linewidth}{0.5pt}\end{center}

\section{Reproducibility}\label{reproducibility}

The full implementation, experiment configurations and result tables are available at \url{https://github.com/ron-42/quantum-kernel-benchmark}. Every figure and table in this paper is generated programmatically from the released result tables, which record one row per configuration and fold; no number was transcribed by hand. Random seeds, hyperparameter grids and library versions are pinned in the release, and the whole study reproduces from raw data with a single command. The hypotheses and analysis choices reported in \mbox{Sec.~\ref{hypothesis-verdicts}} were registered before any experiment was run.

\bibliographystyle{unsrt}
\bibliography{references}

\end{document}